\documentstyle[buckow]{article}

\def\beq{\begin{equation}}
\def\eeq{\end{equation}}

\def\sk{\vskip 1em}
\def\noi{\noindent}

\def\al{\alpha}

\def\ga{\gamma}
\def\th{\Theta}

\def\Ah{\widehat{A}}
\def\Dh{\hat{D}}
\def\phih{\widehat{\phi}}

\def\Fh{\widehat{F}}

\def\CLh{\widehat{\CL}}

\def\d{{\mbox{\tiny{$D$}}}}
\def\D{\delta}

\def\sma#1{\mbox{\footnotesize #1}}

\def\11{\mbox{$1$}}
\def\pa{\partial}
\def\Cal{\cal}

\def\det{{\rm{det}}}
\def\CH{{\cal{H}}}
\def\CL{{\cal{L}}}
\def\CF{{\cal{F}}}

\def\ap{{\alpha^\prime}}
\def\DBI{{\rm{DBI}}}

\def\D3{{\rm{D3}}}

\def\2{{\geq 2}}

\def\e{{\mbox{$\varepsilon$}}}

\def\E{{\mbox{\LARGE $\varepsilon$}}}

\def\Bh{{\hat{B}}}
\def\thE{{\th^{\varepsilon}}}

\def\SW{\mbox{SW}}
\def\bE{{b^\varepsilon}}
\def\gE{{g^\varepsilon}}

\def\Th{\Theta}

\def\U{\Upsilon}
\def\gEs{{g_s^\varepsilon}}
\def\GEs{{G^{{}^{\mbox{\scriptsize{$\varepsilon$}}}}_s}}
\def\CB{{\Cal{B}}}
\def\CE{{\Cal{E}}}
\def\CD{{\Cal{D}}}
\def\vCB{{\CB}}
\def\vCE{{\CE}}
\def\vCH{{\CH}}
\def\CA{{\Cal{A}}}
\def\ccdot{\!\cdot\!}

\def\Hc{{\check{H}}}

\def\q{q}
\def\thth{{{\mbox{\boldmath ${\th}$}}}}
\def\EE{{{\mbox{\boldmath ${\E}$}}}}
\def\1{{-1}}
\def\*{\star}

\newcommand{\eq}{\begin{equation}}
\newcommand{\en}{\end{equation}}
\newcommand{\beeq}{\begin{equation}}
\newcommand{\beqn}{\begin{equation}}
\newcommand{\eeqn}{\end{equation}}
\newcommand{\eneq}{\end{equation}}
\newcommand{\beqr}{\begin{eqnarray}}
\newcommand{\eeqr}{\end{eqnarray}}

\newcommand{\matc}{\begin{array}{c}}
\newcommand{\matcc}{\begin{array}{cc}}
\newcommand{\matccc}{\begin{array}{ccc}}
\newcommand{\matcccc}{\begin{array}{cccc}}
\newcommand{\emat}{\end{array}}

\newcommand{\IH}{\relax{\rm I\kern-.18em H}}
\newcommand{\IR}{\relax{\rm I\kern-.18em R}}
\newcommand{\IK}{\relax{\rm I\kern-.18em K}}
\newcommand{\II}{\hbox{\rm 1\kern-.35em 1}}
\newcommand{\Is}{\relax{\rm 1\kern-.35em 1}}
\begin{document}

\def\titleline{
{\vskip -2.1cm}
Monopoles in Space-Time Noncommutative\\ Born-Infeld
theory
\footnote{Talk presented at the Symposium: {\sl {100 Years Werner 
Heisenberg - Works and Impact}}, workshop 3: {\sl Quantum
Field Theory and Gravitation},  
September 26-30, 2001, Bamberg, Germany. }
}
\def\authors{
Paolo Aschieri}
\def\addresses{Sektion Physik der Ludwig-Maximilians-Universit\"at\\
Theresienstr. 37, D-80333 M\"unchen, Germany

{\tt e-mail aschieri@theorie.physik.uni-muenchen.de} \\
}
\def\abstracttext{
We transform static solutions of space-noncommutative Dirac-Born-Infeld theory
(DBI) into static solutions of space-time noncommutative DBI.
Via Seiberg-Witten map we match this symmetry transformation with a 
corresponding symmetry of commutative DBI. This allows to${}^{\,}$:${}^{\,}$ 1) study new 
BPS type magnetic monopoles, with constant electric and magnetic background
and describe them both in the commutative and in the noncommutative setting;
2) relate by S-duality space-noncommutative magnetic monopoles
to {\sl space}-noncommutative electric monopoles.}
\large
\makefront
\section{Introduction}
Dirichlet branes effective actions can be described by
noncommutative gauge theories, the noncommutativity arising from 
a nonzero constant background NS $B$ field, 
see \cite{Seiberg:1999vs} and references therein.
In fact, when $B\not=0$, the effective physics on the D-brane 
can be described both by a commutative  gauge 
theory $\CL(F+B)$ and by a noncommutative one $\CLh(\Fh)$, where 
\eq
\Fh_{\mu\nu}=\pa_\mu \Ah_\nu - 
\pa_\nu \Ah_\mu -i[ \Ah_\mu\*_{\!\!\!{\textstyle{,}}} \Ah_\nu]
\en 
and  
$\*$ is the Moyal star product;  on coordinates
$[x^\mu\*_{\!\!\!{\textstyle{,}}}x^\nu]=
x^\mu\*x^\nu-x^\nu\*x^\nu=i\th^{\mu\nu}$. The noncommutativity parameter 
$\th$ depends on $B$  and on the metric on the D-brane.   
The commutative/noncommutative descriptions are 
related by Seiberg-Witten map (SW map) \cite{Seiberg:1999vs}. 
Initially  space-noncommutativity has been considered 
($\th^{ij}\not= 0$ i.e. $B_{ij}\not=0$) then  theories also with time 
noncommutativity ($\th^{0i}\not=0$ i.e $B_{0i}\not=0$) have been studied 
\cite{Seiberg:2000gc}.
It turns out that unitarity of noncommutative Yang-Mills theory (NCYM) 
holds only if $\th$ is space-like or light-like 
i.e.$ $ only if the electric and magnetic components of $\th$ (or $B$) are 
perpendicular and if the electric component is not bigger in magnitude
than the magnetic component.
These are precisely the NCYM theories that can be obtained from open 
strings in the decoupling limit 
$\ap\rightarrow 0$ \cite{Aharony:2000gz}. 
In this talk we consider these two  kinds of space-time 
noncommutativity.

In Section 2 we show that for any space-noncommutative static solution 
we can turn 
on time-noncommutativity and obtain a static 
solution with space-time noncommutativity.
This holds in particular for solutions of noncommutative 
Dirac-Born-Infeld-theory (NCDBI).
Via SW map we
obtain the nontrivial action of this symmetry  in 
the corresponding commutative DBI theory and show that 
it is a rotation (boost) between the time component $A_0$ of the gauge 
potential and the worldvolume D-brane coordinates $x^i$. 
This boost, first studied in \cite{Gibbons:1998xz}, is similar to the
target space rotation that relates the linear monopole to the
nonlinear monopole \cite{Moriyama:2000mm}, 
\cite{Gibbons:1998xz}. 
In Section 3 we study BPS solutions of NCDBI and
DBI theories with both electric and magnetic background.
In \cite{Gross:2000wc,Gross:2000ph} solutions to the BPS equations 
of noncommutative electromagnetism (NCEM), and of NCDBI, with just
space-noncommutativity are found.
The solution in \cite{Gross:2000wc} describes 
a smeared monopole connected with a string-like flux tube and is interpreted 
as  a D1-string ending on a D3-brane with constant magnetic field background.
We see that this solution remains a BPS solution also when we 
turn on time-noncommutativity. The corresponding commutative
configuration is also found: It is a new BPS configuration.
Its energy is the energy of the intitial BPS monopole with only 
magnetic field plus the energy of a constant electric field in DBI theory.  
The new BPS configuration describes a monopole plus string in a background
that is both electric and magnetic. The monopole has the fundamental
magneton charge and the string tension is that of a D1-string, this
strongly suggests that we have a D1-D3 brane system 
with both electric and magnetic background. 
The D1-string tension is also 
matched with the string tension of the corresponding space-time 
noncommutative BPS monopole.

Finally in Section 4 we address the issue of duality rotations in
NCDBI and NCEM 
\cite{Gopakumar:2000na,Ganor:2000my}.
At the field theory level duality is present only if $\th$ 
is light-like i.e. the magnetic and electric component of $\th$ are 
perpendicular and equal in magnitude \cite{Aschieri:2001ks,aschieri}. 
For $\th$ space-like as shown in \cite{Gopakumar:2000na} 
we do not have noncommutative gauge theory self-duality and the S-dual 
of space-like NCEM is a noncommutative open string theory decoupled from 
closed strings. A main point here is that under S-duality the magnetic
background is mapped into an electric one, and this background
does not lead to a field theory in the $\ap\rightarrow 1$ limit.
Since $\th$ must be light-like, it may seem that duality rotations
have a very restricted range of application. 
This is not the case because the symmetry we present
in Sections 2 and 3 changes the
background too. It turns out that it is possible to compose duality
rotations with this symmetry, we are thus able to consider
duality rotations with background fixed and arbitrary. 
In particular we briefly discuss the S-dual of the D1-string D3-brane 
configuration of \cite{Gross:2000wc};
it describes an electric monopole plus string in a magnetic
background, possibly a fundamental string ending on a D3-brane in the presence
of a constant magnetic field.

\def\th{\theta}
\section{Gauge theory with space-time noncommutativity}
We use the following notations:
$\Th$ is a generic constant noncommutativity tensor, we have 
$[x^\mu\*_{\!\!\!{\textstyle{,}}}\,x^\nu]=i\Th^{\mu\nu}\,$;
$\th$ is just a space-noncommutativity tensor $\th^{ij}$, 
$\th^{0i}=0\,$;
$\thE$ is a space-time noncommutativity tensor obtained from $\th$
adding electric components, $\thE^{ij}=\th^{ij}$, $\thE^{0i}$.
In three vector notation the electric and magnetic components of $\th$,
respectively $\thE$, are $(0,\thth)$ and $(\EE,\thth)$.
The background fields corresponding to $\Th,\th,\thE$ are 
$B,b,\bE$. 

Consider a noncommutative Lagrangian
$\CLh^\th=\CLh(\Fh,\phih,G,\star_\th)$
where $\phih$ are scalar fields and $G$ is the metric.
The equations of motion (EOM) for $\CLh^\th$ read 
\eq
f_\al(\Fh,\phih,G,\star_\th)=0 \label{space}
\en
where $f_\al$ are functions of the noncommutative fields and their derivatives.
We notice that a static solution of the $\CLh^\th$ EOM (\ref{space})
is also a static solution of 
(\ref{space}) with $\thE$ instead of $\th$, i.e. it is a static
solution of the  $\CLh^\thE$ EOM. Indeed the star products 
$\star_\th$ and
$\star_\thE$ act in the same way on time independent fields.
Moreover the energy and charges of the solution are invariant.
A similar $\th$-$\thE$ symmetry property holds if the fields are
independent 
from a coordinate $x^\mu$ (not necessarily $t$).
This $\th$-$\thE$ symmetry property of static solutions can be used to 
construct moving solutions of a space-noncommutative theory $\CLh^\th$ 
from static solutions of the same theory
$\CLh^\th$.  
Indeed, given  $\th$,  if we turn on an electric component such that 
${\EE}\perp \thth$ and $|\EE | < | {\thth}|$, 
then with a Lorentz boost we can transform this new $\thE$ 
into a  
space-like $\th '$ proportional to the initial $\th$. Rescaling 
$\th'\rightarrow \th$ we thus obtain 
a solution (moving with constant velocity) of the space-noncommutative 
Lagrangian $\CLh^\th$.

We now use SW map and study how the $\th$-$\thE$ 
symmetry acts in the commutative 
theory. We have to consider the two SW maps $\SW^\th$ and $\SW^\thE$. 
In general a static solution 
$\phih$, $\Ah_\mu$ is mapped by $\SW^\th$ and $\SW^\thE$ into two 
different commutative solutions, however if $\Ah_0=0$ then 
$\SW^\th=\SW^\thE$. This can be seen from the index structure of SW map.
In general we have
\beqr
A_\mu &=& \Ah_\mu+\sum_{n>s} 
(\Th^{(n)}\pa^{(n+s)}\Ah^{(n-s)}\Ah)_\mu\nonumber\\[-.6em]
& &\label{SWmap}\\[-.47em]
\phi&=&\phih  +\sum_{n>s} \Th^{(n)}\pa^{(n+s)}\Ah^{(n-s)}\phih\nonumber
\eeqr
where the number of times $n,\;n+s,\;n-s$ that $\Th,\;\pa,\;\Ah$
appear is dictated by dimensional analysis. In (\ref{SWmap}) 
we do not specify which $\pa$ acts on which $\Ah$ and we do not
specify the coefficients of each addend. Because of the index structure
we notice that $\Th^{0i}$ never enters (\ref{SWmap}) 
if $\phih,~\Ah$
are time independent and $\Ah_0=0$. 
The commutative fields $\phi,\;A_i$ corresponding
to $\phih,~\Ah_i$ ($i\not=0$) are solution of both 
$\CL^{\th}$ and $\CL^{\thE}$. Here $\CL^{\th}$ and $\CL^{\thE}$ are 
the commutative Lagrangians
corresponding to $\CLh^\th$ and $\CLh^\thE$ via SW map.
In the case of the DBI Lagrangian with a scalar field $\phi$,
$\CL^{\th}$ and $\CL^{\thE}$ 
reads 
\beq
\CL_\DBI(F+b,\phi,g,g_s)=
\frac{-1}{\ap^2 g_s}\sqrt{-{\rm{det}}(g+\ap (F+b)+\ap^2\pa\phi_a\pa\phi^a)}
\eeq
and 
$\CL_\DBI(F+\bE,\phi,\gE, g_s)$. We should write $\gEs$ instead of
$g_s$ in this last expression, however we can
rescale $G_s$ and thus impose the invariance of the 
closed string coupling constant $g_s$.  
The relation between closed and open string parameters is given by
(see \cite{Seiberg:1999vs}): $(g+\ap B)^{-1}=G^{-1}+{\Theta}/{\ap}\,$,
$\,G_s=g_s 
\sqrt{{{\rm{det}}G}{{}^{\,}} {{\rm{det}}{(g+\ap B)}^{-1}}}
$.
\def\gEs{g_s}

In order to have a more explicit formulation of the $\th$-$\thE$
symmetry, from now on 
we set the noncommutative 
open string metric $G=\eta=diag(-1,1,1,1)$ and we consider (rigid) coordinate 
transformations $x\rightarrow x''$ and $
x\rightarrow x'$ that respectively orthonormalize the closed string
metrics $\gE$ and $g$,
while preserving time independence of the transformed $\phi\,,\,A$ 
fields. The result is that
if $\phi'\,,\,\CA'_i\,,\CA'_0=0$ is a static solution of $\CL_\DBI$ then 
$\phi''\,,\,\CA_{\mu}''$ is a new static solution  \cite{aschieri}. Here
$\CA'$ is the gauge potential of $\CF'=F'+b'$ in the $x'$ basis and
\eq
\phi''(x'')=\phi'(x')~\,,~~~\CA_i''(x'')=\U^j_{~i} \CA'_j(x')~
\,,~~~\CA_0''(x'')=e''x''_2
 ~~ \label{Newsol}
\en
with $x''=\U^{-1}x'$; the nonvanishing components of $\U$ are
$$
\U_{00}=\U_{22}=\sqrt{1-\ap^2{e''}^2}~,~~
\U_{11}=\U_{33}=1~,~~ \U_{01}=\ap^2{e''}^{\,}b'~~~$$
and we have turned on time-noncommutativity just in the $x''_2$ direction.

Which is the symmetry of commutative $\CL_\DBI$ 
that underlies this family of solutions?
We split $\CF'$ in its electric field $\CE'$ and magnetic 
induction $\CB'$ components.  
We then consider the Legendre transformation 
of $\CL_\DBI$ 
\eq
\Hc(\vCE',\vCH',\phi')=\frac{1}{g_s}\vCB' \ccdot \vCH' + \CL_\DBI~~,~~~
{\mbox{where}}~~~~
\vCH'_i= -g_s\frac{\pa\CL_\DBI}{\pa \vCB'^i} ~~\label{leg}
\en
For time independent fields we have that  the EOM imply
$\vCH'=-{\nabla'}\chi'$
and  $\vCE'=-{\nabla'}\psi'$, ($\psi'=-\CA'_0$). 
As shown in  \cite{Gibbons:1998xz} it follows that  $\Hc(\psi',\chi',\phi')$
is the action  of a space-like 3-brane immersed in a 
target space of coordinates $X'^A=\{\ap\psi',\ap\chi',\ap\phi',{x'}^{\,i}\}$ 
and metric $\eta={diag}(-1,-1,1,1,1,1)$
\eq
\int\!\!d^3x' ~\Hc~~=\;~\frac{-1}{g_s\ap^2}\int\!\!d^3x'~ 
\sqrt{\det 
\left(\eta_{AB}\frac{\pa X'^A}{\pa x'^{\,i}}\frac{\pa X'^B}{\pa x'^{\,j}}\right)\,}
\label{H}~~.
\en
It is the $SO(2,4)$  symmetry \cite{Gibbons:1998xz} 
of this static gauge action 
that is relevant in our context: Consider the Lorentz transformation 
$Y^A=\Lambda^A_{~B}X'^B$ (where $X'^{\,i}=x'^{\,i}$) and express $Y^A$ as 
$Y^A=Y^A(y^i)$  (where $Y^i=y^i$) so that we are still in static gauge;
the action (\ref{H}) is invariant under $X'^A(x'^{\,i})\rightarrow Y^A(y^i)$.
In particular a boost in the $\ap\psi'\,,\;x'_2$ 
plane with velocity $\beta=- \ap e''$ gives (\ref{Newsol})
\cite{aschieri}.

\section{BPS solutions for (NC)DBI with a scalar field}
A BPS solution of DBI theory in the $x$-reference system with metric
$g$ and background $b=\th/(\ap^2+\th^2),$ $\th=-\th^{12},$ all others
$\th^{\mu\nu}=0$
is given by
\eq
{\phi}=\frac{{\,}-\th{\;}}{\ap^2}\,{x}^3 - \frac{1}{2 r}~~,~~~
\CB_i=-\pa_i\phi~~;~~~
r^2=g_{ij}x^i x^j
=({x}^3)^2+
\frac{(x^1)^2+(x^2)^2}{1+\th^2/\ap^2} \label{bpssol}  
\en
where $\frac{1}{2 r}=\frac{q_m}{4\pi r}$ with $q_m$ the magneton 
charge $q_m=2\pi$.
This solution describes a D1-string ending on a D3-brane.
Because of the magnetic backgound field $b$ on the brane the 
string is not perpendicular to the brane, in (\ref{bpssol})
the string is vertical and the brane is tilted w.r.t. the horizontal direction.
The magnetic force acting 
on the end of the D1-string is compensated by the tension of the
D1-string.  

A BPS solution of NCDBI with  space-like noncommutativity
given by $\th^{12}=-\th$, (all others $\th^{\mu\nu}=0$) has been 
studied in \cite{Gross:2000wc}. It is a static solution with $\Ah_0=0$.
This noncommutative BPS solution is caracterized by a noncommutative
string tension and a magneton charge. It is expected to correspond, via SW
map (and the target space rotation relating the linear monopole to the
nonlinear one \cite{Moriyama:2000mm}), to (\ref{bpssol}).
A first evidence is the correspondence between the tension of the
noncommutative string and that of the D1-string \cite{Gross:2000wc}, then 
in \cite{Gross:2000ph} the spectrum of small fluctuations around (a limit of) 
this solution is studied and found in agreement with the expectations 
from string theory. 

We now discuss what happens to these commutative/noncommutative BPS
solutions when we apply the
$\th$-$\thE$ symmetry. The result \cite{aschieri} is that we obtain
two new BPS solutions that describe a D1-string ending on a
D3-brane with both an electric and magnetic background.
The noncommutative one is obtained simply writing $\thE$ instead of
$\th$ (and rescaling the open string coupling constant
$G_s\rightarrow \GEs$ since we keep $g_s$ invariant), it still
satisfies the noncommutative BPS equations 
$\Bh_j=-\Dh_j\phih$ with $\Dh_j\phih=\pa_j\phih-i[\Ah_j\*_{\!\!\!{\textstyle{,}}}\phih]$.
The commutative solution is most easily written in the orthonormal  reference system $x''$:
\beqr
\CB''_2&=&-\gamma\,\pa''_2\phi''\nonumber\\
\CB''_\q&=&-\gamma^{-1}\,\pa''_\q\phi''\label{BPSpp2}~~~~~~~~~~~~~~
\sma{$\q=1,3$}\\
\CE''_2&=&e''~~~,~~~~\CE''_\q=0~~\nonumber
\eeqr
with $\ga^{-1}=\sqrt{1-\ap^2 e''^2}$ and
\eq
\phi''=-\gamma\,b'' x''^{\,3}-\frac{1}{2 \, R}
~,~\label{solpp}
~~R^2\equiv (x''^{\, 1})^{2}+
\gamma^{-2} (x''^{\, 2})^{2} + (x''^{\, 3})^{2}\nonumber
\en
Eq.s (\ref{BPSpp2}) are
obtained from the ($x'$-reference system) BPS equations
$\CB'_i=-\pa_i\phi'$ 
via the boost $\U$, cf. (\ref{Newsol}). The nonvanishing components of
$b''_{\mu\nu}$ and $\th^{\mu\nu}$ are 
$e''=-b''_{02}=
\frac{-{{\mbox{\large $\varepsilon$}}}}{\ap\sqrt{\ap^2+\th^2}},\,
b''=b''_{12}={\frac{\th}{\ap^2}}\sqrt{1-\frac{\e^2}{\ap^2+\th^2}},\,
\th=-\th^{12},\,\E=\thE^{02}$.
A solution of (\ref{BPSpp2}) has  
an energy
\eq
\Sigma''=
\Sigma'+\frac{1}{g_s}\int\!d^3x''\,\CE''_i\CD''^{\,i}~\label{93}
\en
where $\Sigma'$ is the energy of the corresponding solution of 
$\CB'_i=\pa_i\phi'$. 
We see that the energy is of BPS type, indeed it is the sum  of the old BPS
energy $\Sigma'$ plus the topological charge $Z''_e=\frac{1}{g_s}\int\!d^3x''
\,\CE''_i\CD''^{\,i}=-\frac{1}{g_s}\int\!d^3x''\,\pa_i\psi''\CD''^{\,i}\,$.
The explicit value of $\CD''^{\,i}\equiv g_s\frac{\pa\CL_\DBI}{\pa\CE''_i}$
is
$
\CD''^{\,2}= e''\gamma,~ \CD''^{\,1}=\CD''^{\,3}=0~.
$
We can also write
$\Sigma''=\frac{1}{\ap^2g_s}\int\!\!d^3x''+|Z''_m|+
\frac{\ga-1}{\ap^2g_s}\int\!\!d^3x'' $,
and recognize the brane tension, the topological charge
$Z''_m=\int\!d^3x''\,\pa_i\phi''\CB''^{\,i}\,
=\int\!d^3x'\,\pa_i\phi'\CB'^{\,i}\,=Z'_m$
and the energy of just the electric field $e''$ in DBI theory.
We also have that the magnetic charge and the string tension associated to  
solution (\ref{BPSpp2}),(\ref{solpp})  are  those of a D1-string 
as we expect from a BPS state.
Notice that the shape of the funnel representing this D1-string is
no more symmetric in the $x''_1,x''_2$ directions. A section 
determined by $\phi''=\,${\it const} , $x_3''=0$ is an ellipsis in the 
$x_1'',x_2''$ plane.
The ratio between the  ellipsis axes is given by
$\gamma$.  One can project the D1-string 
on the D3-brane and consider the tension of this projected string: 
It matches the tension associated to the
corresponding noncommutative BPS solution. 
\sk

\section{Dual string-brane configuration}

If we duality rotate the D1-D3 brane configuration (\ref{BPSpp2}),(\ref{solpp}) 
we obtain a soliton solution that describes a fundamental 
string ending on a D3-brane with electric and magnetic background.
Under a $\pi/2$ duality rotation we have (we set $2\pi=1$, recall also
$g''_{\mu\nu}=\eta_{\mu\nu}$)  
\eq
g_s^{\d}=\frac{1}{g_s}~~~,~~~~~
g{''{}^\d}_{\!\!\!\!\!\!\mu\nu}=\frac{1}{g_s}\eta_{\mu\nu}~~~,~~~~~
\phi''_\d =\left(\frac{1}{g_s}\right)^\frac{1}{2}\,\phi''~~
\label{phiDual}
\en
the dual of solution (\ref{BPSpp2}),(\ref{solpp}) is given by (\ref{phiDual}) and 
\eq
\CA''^\d_0=-\left(\frac{1}{g_s}\right)^\frac{1}{2}\phi''_\d~~,~~~
\CA''^\d_1=-\frac{1}{g_s}\ga\,e''x''^3~~,~~~\CA''^\d_2=\CA''^\d_3=0
\label{ASWm1}~~~.
\en 
Is there a noncommutative field theory description
of the F1-D3 system?
Since NCDBI and its $\ap\rightarrow 1$ limit, NCEM,
admit duality rotations only if
$\th$ is light-like,
it seems that we have a F1-D3 system only if we consider
a light-like background. This light-like condition 
may appear a strong constraint. Actually, using the $\th$-$\thE$
symmetry we are {\sl not} bound to consider only this restrictive case of 
light-like background. Indeed for 
any space-noncommutative static solution we can turn on time-noncommutativity 
and obtain a static solution with light-like noncommutativity. We can then 
apply a duality rotation, switch off the time-noncommutativity and thus 
obtain a new solution of the original pure space-noncommutative theory.

In particular in order to obtain the duality rotated configuration 
of the one described in \cite{Gross:2000wc}, we consider the
corresponding commutative configuration (\ref{bpssol}), that in the
$x'$ orthonormal frame reads $\,{\cal{B}}'_i=-\pa'_i\phi'$,
$\,\phi'=-{1\over \ap}\th x'^{\,3}-\frac{1}{2 \, r'}
\;,~{r'}^2\equiv (x'^{\, 1})^{2}+(x'^{\, 2})^{2} + (x'^{\, 3})^{2}$.
In order to have a light-like background
we turn on a constant electric field keeping here fixed $G_s$ besides
the open string metric $G=\eta$ (therefore here $g_s\rightarrow 
g_s^\varepsilon\not= g_s$).
We then obtain (\ref{BPSpp2}) and (\ref{solpp}) with $e''=-b''$ (i.e. 
$\th=\E$).
Next we  duality rotate this solution and obtain
(\ref{phiDual}) and (\ref{ASWm1}) with
$g_s^{\varepsilon\d},g^\varepsilon_s$ 
instead of $g_s^\d, g_s$. 
Finally we go back to the original $x$-reference system
(the $x\rightarrow x''$ coordinate
transformation commutes with duality rotations),
we apply SW map and arrive at the noncommutative fields 
$\Ah^\d,\phih_\d$ with open string coulping constant, metric 
and light-like noncommutativity given by
\eq
 G_s^\d={1\over G_s}~,~~G_{\mu\nu}^\d={1\over G_s}\eta_{\mu\nu}~,~~
\thE^{\mu\nu}_{\!\!\!\!\d}=
{1\over 2} {G_s\,}\epsilon^{\mu\nu\rho\sigma}\thE_{\!\!\!\rho\sigma}\,.
\en 
The noncommutative fields  $\Ah^\d,\phih_\d$ correspond to a fundamental string
ending on a D3-brane with light-like background.
These 
fields solve also 
$\CLh_\DBI(\Fh,\phih,G^\d,G^\d_s,\*_{\th_\d})$
where 
$\th_\d$ ($\th_\d^{13}=G_s\th,$ all others $\th_\d^{\mu\nu}=0$) 
is just the space part of $\thE_{\!\!\!\!\d}\,$; they describe an
electric monopole with a string attached.
Since the noncommutative string tension and charges are invariant under
$\th_\d^\varepsilon\rightarrow \th_\d$, the $\Ah^\d,\phih_\d$ fields 
are a good candidate to describe an F-string ending on a
D3-brane with constant magnetic background.

\sk
\sk
\noi{\bf{Acknowledgements}}\\
{{I wish to thank the organizers  and 
Alexander von Humboldt-Stiftung for the 
stimulating atmosphere of the conference.
  This work has been supported by 
a Marie Curie Fellowship of the European Community programme IHP under
contract number MCFI-2000-01982}}



\begin{thebibliography}{77}


\bibitem{Seiberg:1999vs}
N.~Seiberg and E.~Witten,
JHEP{\bf 9909} (1999) 032
[hep-th/9908142].


\bibitem{Seiberg:2000gc}
N.~Seiberg, L.~Susskind and N.~Toumbas,
JHEP{\bf 0006} (2000) 044
[hep-th/0005015].\\
J.~L.~Barbon and E.~Rabinovici,
Phys.\ Lett.\ B {\bf 486} (2000) 202
[hep-th/0005073].\\
J.~Gomis and T.~Mehen,
Nucl.\ Phys.\ B {\bf 591} (2000) 265
[hep-th/0005129].



\bibitem{Aharony:2000gz}
O.~Aharony, J.~Gomis and T.~Mehen,
JHEP{\bf 0009} (2000) 023
[hep-th/0006236].



\bibitem{Gibbons:1998xz}
G.~W.~Gibbons,
Nucl.\ Phys.\ B {\bf 514} (1998) 603
[hep-th/9709027].

\bibitem{Moriyama:2000mm}
K.~Hashimoto and T.~Hirayama,
Nucl.\ Phys.\ B {\bf 587} (2000) 207
[hep-th/0002090].\\
S.~Moriyama,
Phys.\ Lett.\ B {\bf 485} (2000) 278
[hep-th/0003231].\\
S.~Moriyama,
JHEP {\bf 0008} (2000) 014
[hep-th/0006056].\\
K.~Hashimoto, T.~Hirayama and S.~Moriyama,
JHEP {\bf 0011} (2000) 014
[hep-th/0010026].

\bibitem{Gross:2000wc}
D.~J.~Gross and N.~A.~Nekrasov,
JHEP {\bf 0007} (2000) 034
[hep-th/0005204].

\bibitem{Gross:2000ph}
D.~J.~Gross and N.~A.~Nekrasov,
JHEP {\bf 0010} (2000) 021
[hep-th/0007204].

\bibitem{Gopakumar:2000na}
R.~Gopakumar, J.~Maldacena, S.~Minwalla and A.~Strominger,
JHEP{\bf 0006} (2000) 036
[hep-th/0005048].


\bibitem{Ganor:2000my}
O.~J.~Ganor, G.~Rajesh and S.~Sethi,
Phys.\ Rev.\ D {\bf 62} (2000) 125008
[hep-th/0005046].\\
S.~Rey and R.~von Unge,
Phys.\ Lett.\ B {\bf 499} (2001) 215
[hep-th/0007089].\\
J.~X.~Lu, S.~Roy and H.~Singh,
Nucl.\ Phys.\ B {\bf 595} (2001) 298
[hep-th/0007168].


\bibitem{Aschieri:2001ks}
P.~Aschieri,
Mod.\ Phys.\ Lett.\ A {\bf 16} (2001) 163
[hep-th/0103150].

\bibitem{aschieri}
P.~Aschieri,
Nucl.\ Phys.\ B {\bf 617} (2001) 321  [hep-th/0106281].


\end{thebibliography}
\end{document}